# Back-to-back Converter Control of Grid-connected Wind Turbine to Mitigate Voltage Drop Caused by Faults

Fattah Hassanzadeh
Electrical and Instrument Engineering Department
Mapna Turbine Engineering and Manufacturing Company
Karaj, Iran
Hassanzadeh@mapnaturbine.com

Hossein Sangrody, *Student Member, IEEE*
Electrical and Computer Engineering
State University of New York at Binghamton
Binghamton, USA
habdoll1@binghamton.edu

Amin Hajizadeh, *Senior Member, IEEE*
Department of Energy Technology
Aalborg Universitet Esbjerg
Esbjerg, Denmark
aha@et.aau.dk

Shahrokh akhlaghi, *Student Member, IEEE*
Electrical and Computer Engineering
State University of New York at Binghamton
Binghamton, USA
Sakhlag1@binghamton.edu

*Abstract*—Power electronic converters enable wind turbines, operating at variable speed, to generate electricity more efficiently. Among variable speed operating turbine generators, permanent magnetic synchronous generator (PMSG) has got more attentions due to low cost and maintenance requirements. In addition, the converter in a wind turbine with PMSG decouples the turbine from the power grid, which favors them for grid codes. In this paper, the performance of back-to-back (B2B) converter control of a wind turbine system with PMSG is investigated on a faulty grid. The switching strategy of the grid side converter is designed to improve voltage drop caused by the fault in the grid while maximum available active power of wind turbine system is injected to the grid and the DC link voltage in the converter is regulated. The methodology of the converter control is elaborated in details and its performance on a sample faulty grid is assessed through simulation.

*Index Terms*-- Back-to-back (B2B) converter, direct-in-line wind turbine, permanent magnetic synchronous generator (PMSG), voltage drop, wind turbine control.

## I. Introduction

Along with environmental benefits of renewable energies, the advent of new technologies in the operation and control of renewable energy sources and the increasing demand for high quality and consistent supply result in more attentions to this type of energy sources [1-4]. Besides optimal operation of a power system in normal condition, controlling the system in faulty condition is one of the most challenging concerns [5-7]. Power quality is one of the most significant subjects in utilizing the distributed generation (DG) [8-9]. In addition to frequency and active power, voltage and reactive power must be bounded and controlled in predefined ranges [10-11]. To achieve the goal, the generators must be controlled in a properly, so that the voltage drop and voltage rising are avoided at peak time or low demand, respectively [12-15].

One of the most important advantages of DGs, other than active power injection and local load supply, is reactive power injection at the point of common coupling (PCC) [16-17]. By controlling the reactive power of a DG, voltage profile and power quality can be improved in different operational modes of the grid especially during faults. Wind turbine generation is one of the most common sources used in distributed generation systems for this purpose.

Constant speed wind generators which were more popular in the eighties are less efficient in compared with the recent wind turbine systems in which power electronic technology has helped improve the efficiency by implementing variable speed wind generators in power generation [18]. As shown in Fig. 1, wind turbine generators can be classified into two groups of constant and variable speeds in which variable speed generators are also divided into three types. Based on the operating and controlling methodologies, topology of connection to the grid, and generator types of wind turbine system, they can be classified into four types in total [19].

In the first type, turbine generator is an asynchronous squirrel cage induction generator (SCIG) which operates at the constant speed and the rotor of the turbine is connected to the generator shaft with a fixed-ratio gearbox. Simplicity is the main advantage of this type of generators; however, drawing fluctuating reactive power and low efficiency are the main disadvantages.

Type B, C, and D, are wind turbines operating at variable speeds. Type B corresponds to partial variable speed turbine which is directly connected to the network by a wound rotor induction generator (WRIG) with variable rotor resistance. Type C includes doubly-fed induction generator (DFIG) which is directly connected to the network and controlled by a partial-scale power converter. Although DFIG has better performance and efficiency than type A and type B, it requires maintenance for the use of brushes in slip rings of the generator. Finally, in type D, a synchronous generator is directly connected to the power system via a full-scale power converter. The synchronous generators for this case are commonly WRSG, or permanent magnet synchronous generator (PMSG) [20]. This type of wind turbine can operate at low speed which eliminates the need for gear box and suitable for low maintenance usage, especially for offshore application. High efficiency of type C and low maintenance of type D is the reasons for wide applications of these types of the wind turbine in recent years.

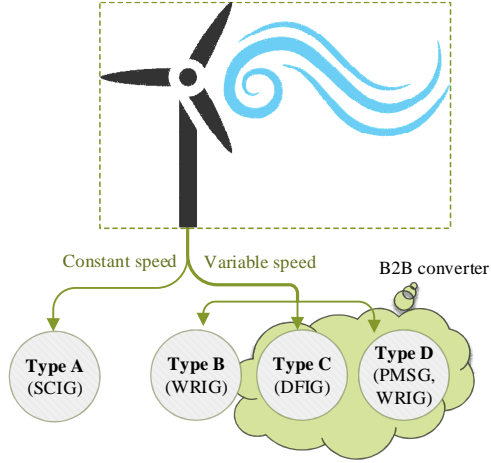

Fig. 1. Wind turbine types

As shown in Fig. 1, both DFIG and PMSG (type C and type D, respectively) use AC/DC/AC (back-to-back (B2B)) converter with low pass filter which allows high efficient performance and reactive power compensation.

Several studies have been dedicated to B2B converters control. Khatir et al. in [21] modeled the B2B converter based on state equations and its dynamic performance during symmetric/non-symmetric faults, and step changes in the active and reactive are investigated. In [22-24] scalar and vector drive control systems are elaborated in details. Huang et al. in [25] presented a wind turbine power generation system using a B2B converter with sensor-less vector control strategy. In [26], B2B converter in a wind turbine system is controlled by direct-current vector control method, where the control objectives of the system are investigated under constant and variable wind speeds.

Although DFIG has a high efficient operation, the control system requires a complex scheme when a fault occurs in the power system. On the other hand, since in type D, the PMSG and the wind turbine is decoupled from the power grid, this configuration is favored in the faulty system [27].

In this paper, the performance of the B2B converter controller in type D of the wind turbine with PMSG is assessed on a faulty power system. The controller of the B2B converter is applied in a wind turbine system with the multipurpose of maximum active power injection to power grids, to control reactive power to compensate grid voltage at PCC [28-29], and also to regulate the DC link voltage at a desired value. In this work, both grid side and generator side converters are controlled by applying vector control and pulse width modulation (PWM) switching strategy. The controlled system is simulated in MATLAB® software.

The rest of paper is organized as follows. In section II, system modeling of the wind turbine and generator is formulated. In section III, the control of B2B converter including generator and grid side controller is elaborated in details. Section IV represents the simulation results of assessing the performance of the turbine controller on both normal and fault conditions of the grid. Finally, the conclusion is drawn in section V.

## II. WIND TURBINE MODELLING

### A. Wind Turbine

Considering wind speed, the power generated by the wind turbine is calculated through the following equation [30]:

$$P_w = \frac{1}{2} \rho A C_p(\beta,\lambda) V_m^3 \quad (1)$$

$$\lambda = R W_m / V_m \quad (2)$$

Where $\rho$ and $V_m$ are the wind density (which typically is 1.225 kg/m³ at sea level for the normal temperature of 15°C and an atmospheric pressure of 101.325 kPa) and wind speed respectively. Also, $A$ is the swept area by the turbine, $\beta$ is the blade pitch angle, $\lambda$ is speed rate, $R$ is the radius, $W_m$ is the rotational velocity (rad/sec), and $C_p$ is power coefficient which depends on the turbine design.

### B. Generator mode

The dynamic model of a PMSG in the $dq0$ reference framework is represented by equation (3).

$$\begin{bmatrix} v_{sd} \\ v_{sq} \end{bmatrix} = -R_s \begin{bmatrix} i_{sd} \\ i_{sq} \end{bmatrix} - \frac{d}{dt}\begin{bmatrix} \psi_{sd} \\ \psi_{sq} \end{bmatrix} + w_e \begin{bmatrix} 0 & -1 \\ 1 & 0 \end{bmatrix}\begin{bmatrix} \psi_{sd} \\ \psi_{sq} \end{bmatrix} \quad (3)$$

Where $R_s$ is the stator winding resistance, $w_e$ the rotational velocity of the generator, $v_{sd}$, $v_{sq}$, $i_{sd}$, $i_{sq}$, $\psi_{sd}$, and $\psi_{sq}$ are stator voltage, current, and flux components in the dq0 reference framework, respectively. If the $d$ axis lines up with the rotor flux, the stator flux will be calculated as follows [26].

$$\begin{bmatrix} \psi_{sd} \\ \psi_{sq} \end{bmatrix} = \begin{bmatrix} L_{ls} + L_{dm} & 0 \\ 0 & L_{ls} + L_{qm} \end{bmatrix}\begin{bmatrix} i_{sd} \\ i_{sq} \end{bmatrix} + \begin{bmatrix} \psi_f \\ 0 \end{bmatrix} \quad (4)$$

Where $L_{ls}$ is the leakage inductance of the stator windings, $\psi_f$ is the rotor flux generated by the permanent magnet, and $L_{dm}$ and $L_{qm}$ are mutual inductances of the stator and rotor in $d$ and $q$ axis, respectively. In addition, the electromagnetic torque is derived in the $dq0$ reference framework by the following equation [26].

$$\tau_m = \rho(\psi_{sd}i_{sd} - \psi_{sq}i_{sq}) = \rho(\psi_f i_{sq} + (L_d - L_q)i_{sd}i_{sq}) \quad (5)$$

Where, $L_d = L_{ls} + L_{dm}$ and $L_q = L_{ls} + L_{qm}$. In the steady state mode, equation (3) turns into the following equation:

$$\begin{bmatrix} v_{sd} \\ v_{sq} \end{bmatrix} = \begin{bmatrix} -R_s & -w_e L_q \\ w_e L_d & -R_s \end{bmatrix}\begin{bmatrix} i_{sd} \\ i_{sq} \end{bmatrix} + \begin{bmatrix} 0 \\ w_e \psi_f \end{bmatrix} \quad (6)$$

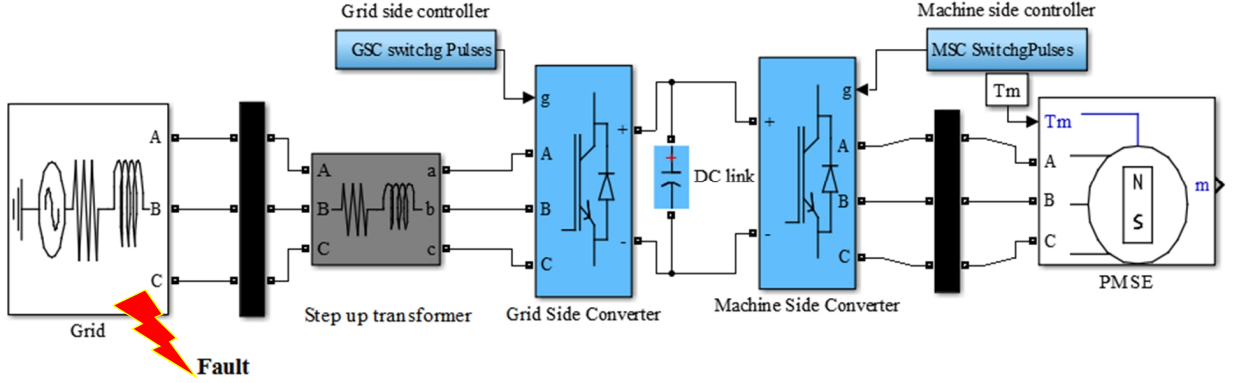

Fig. 2. Schematic of the simulated system

## III. CONTROL OF B2B CONVERTER

Fig. 2 displays the schematic of the simulated system which includes schematic of the wind turbine which is directly connected to a faulty power grid through a B2B converter, a step-up transformer, and a low pass filter. In this configuration, the wind turbine generator is a PMSG and the B2B converters include a machine-side converter (MSC) and a grid side converter (GSC) with a DC voltage line between them converter [31]. In the wind turbine system, a three-phase full wave rectifier and a three-phase three-level voltage source inverter with gate bipolar transistors (IGBTs) are used for MSC and GSC, respectively.

There exist different control mythologies for controlling switches based converters such as PMW, model predictive-based control (MPC), and zero voltage switch-PWM (ZVS-PWM) [32-35]. In this paper, the PWM modulation strategy is applied to generate gate pulses of the IGBTs. In this method, by using high-frequency carrier signal, the input reference AC signal is modulated to high-order frequency and the destructive low-frequency harmonics are diminished [31]. In [33] an advanced control strategy was proposed that can efficiently and reliably address uncertainties and unknown disturbances.

### A. Control of machine-side converter

In the controlling system of the turbine, active power control and also tracking its maximum available value are considered in control objectives of the generator side converter [36-37]. Fig. 3 shows control loop of the MSC. In this figure, the $q$ axis loop is devoted to the generator speed and torque control, and the $d$ axis loop is applied for the other control objectives.

As illustrated in Fig. 3, in the $d$ axis loop, the $d$ axis current of the generator ($i_{sd}$) and the reference current ($i_{sd}*$) are compared and subtracted which provides, the reference voltage signal of the d axis ($v_{sd}'$) through the PI controller.

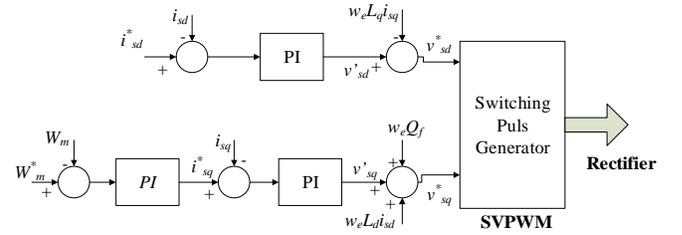

Fig. 3. Control method of MSC

To have a stabilized power, $i_q$ and $w_m$ are correlated and any change in $w_m$ will alter $i_q$. Therefore, through a PI controller, the $q$ axis reference current ($i_{sq}*$) is provided by comparing generator nominal rotational speed and its instantaneous speed ($w_m$). Afterward, by comparing the $q$ axis currents of the generator stator with its reference signal and using the PI controller, the reference voltage of the $q$ axis ($v_{sq}'$) is achieved. Finally, the decoupled voltages, $v_{sd}*$ and $v_{sq}*$, are derived as shown in (7):

$$\begin{cases} v_{sd}^* = v_{sd}' - w_e L_q i_{sq} \\ v_{sq}^* = v_{sq}' - w_e L_d i_{sd} + w_e \psi_f \end{cases} \quad (7)$$

### B. Control of grid-side converter

The control objectives of this converter are:

- To control and regulate the DC link voltage at a predefined reference value
- To control injected reactive power to the grid at a predefined reference value.

Fig. 4 shows the control method of the GSC. As shown, the $q$ axis loop is applied to control the reactive power injection, while the $d$ axis loop is used for controlling the DC link voltage. GSC is required be able to stabilize the DC link voltage within a permissible working range of the converter. In addition, considering the fact that there is a direct relation between the DC link voltage, the active power, and the $d$ axis current of the grid side converter, using a PI controller, it is possible to consider the output signal of the DC link voltage controller as a reference signal for the $d$ axis current. As shown in Fig. 4, by comparing $d$ axis current with its reference signal and through a PI controller, the reference voltage signal for $d$ axis ($v_d'$) is derived.

Reactive power control of the inverter is performed by the $q$ axis control loop so that whenever the voltage drops due to the transient faults on the grid side or when any overvoltage is observed at PCC, the GSC converter should inject a suitable corresponding reactive power to compensate variations of the voltage at PCC. Therefore, as depicted in Fig. 4 for the $q$ axis, by comparing the transformed reactive power ($Q$) with its reference value ($Q_{ref}$) and through a PI controller, the reference for $q$ axis current is provided. Finally, the decoupled voltages, $v_{dl}*$ and $v_{ql}*$, are derived as shown in (8).

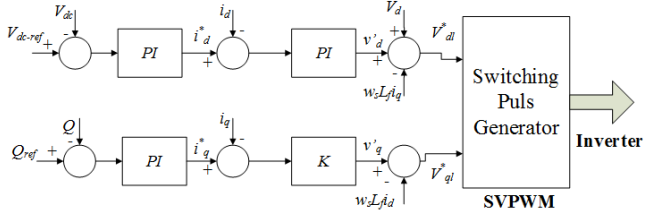

Fig. 4. Control method of GSC

$$\begin{cases} v_{dl}^* = v_d' - w_s L_f i_q + v_d \\ v_{ql}^* = v_q' - w_s L_f i_d \end{cases} \quad (8)$$

## IV. SIMULATION RESULTS

In this section, the performance of B2B controller in the wind turbine system connected to a weak grid at the PCC is simulated in MATLAB®. The performance of the controller is shown for both normal and fault conditions. As mentioned earlier, the main control objectives of the MSC and GSC converters are to inject maximum available active power to the grid, to stabilize DC link voltage when supplying the local loads, and to control the reactive power during a fault or any other circumstances that may affect the voltage at PCC. In this study, the system is investigated in both normal and fault conditions.

### A. Power system under normal condition

In this condition, it is assumed that the wind turbine system works in its normal conditions. So, wind turbine system injects its generated active power to the grid via converters. Fig. 5 shows the DC link voltage in the B2B converter. As illustrated, the DC link voltage is regulated at its desired and constant value. Also, the same results can be seen for the active and reactive powers in Figs. 6 and 7, respectively. As shown in these figures, both powers are stabilized in their steady state. In Fig. 8, the turbine current injected to the grid is depicted.

### B. Power system under fault condition

In this section, as illustrated in Fig. 2, it is assumed that the wind turbine connected to a power grid which is under a fault condition. Such a fault might happen anywhere in the grid; however, there is a consequence voltage drop at PCC. For the simulation, it is assumed that a three-phase fault in the power grid during 0.5 second to 1.2 seconds.

As shown in Fig. 9, when the wind turbine is not connected to the grid, the voltage drops, significantly. However, when the wind turbine is connected to the grid, the voltage drop at PCC is reduced by the wind turbine which injects reactive power to the grid. In other words, the controller of the B2B controller compensates the voltage drop by injecting reactive power while it still fulfills the other aforementioned objectives.

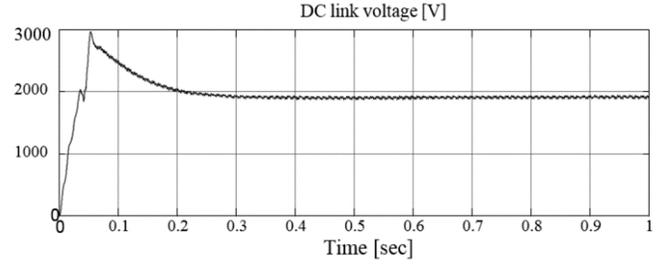

Fig. 5. The DC link voltage in the normal condition

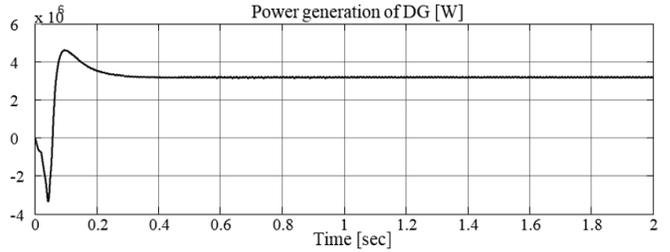

Fig. 6. Active power injected by wind turbine to the grid in the normal condition

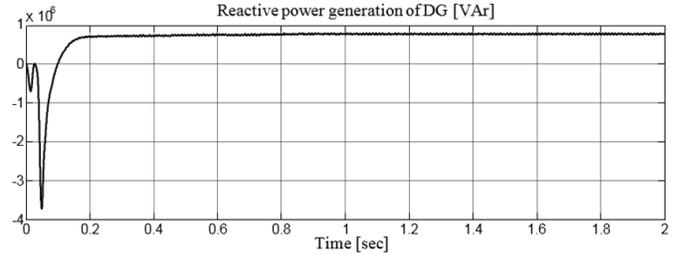

Fig. 7. Reactive power injected by wind turbine to the grid in the normal condition

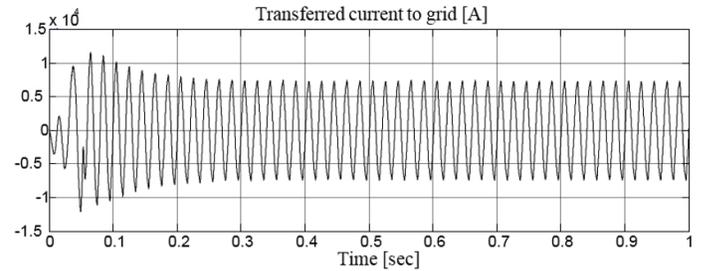

Fig. 8. Injected current to the grid in the normal condition

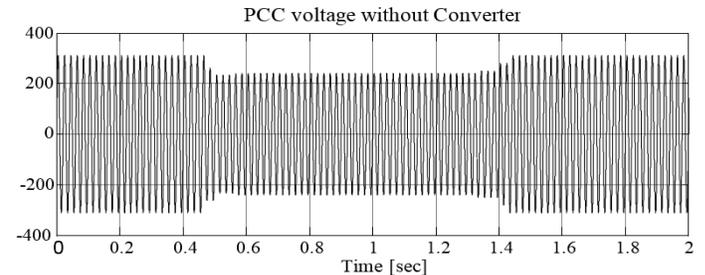

Fig. 9. Voltage at the PCC point without the presence of wind turbine during the fault

Figs. 10, 11, and 12 show the fault current, the active and reactive powers injected into the grid, respectively. As indicated in Fig. 10, the active power injected by the wind turbine shows oscillations during the fault occurrence and fault removal. However, it is stabilized at a constant and in fact its maximum available value. Fig. 12 also illustrates the high performance of converter controller in dealing with voltage drop during the fault. As shown, during the fault time, the reactive power is increased considerably to help voltage drop compensation.

The result of the controller's performance in imitating the voltage drop at PCC is illustrated in Fig. 13. Comparing Fig. 9 (voltage drop without the presence of the wind turbine) and Fig. 13 (voltage drop in the presence of the wind turbine) gives a clearer insight into the controller's performance. In other words, at the presence of the wind turbine, the voltage drop depicted in Fig. 9 is reduced up to 70% because the injected reactive power has increased considerably. Fig. 14 also depicts the injected current by the inverter which increases during the fault time.

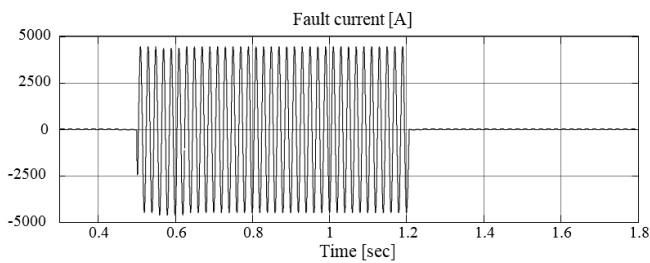
Fig. 10. Fault current at PCC

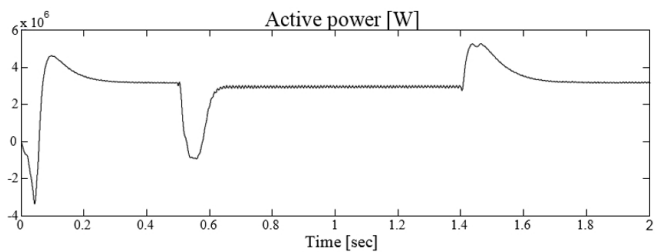
Fig. 11. Injected active power by the wind turbine during the fault

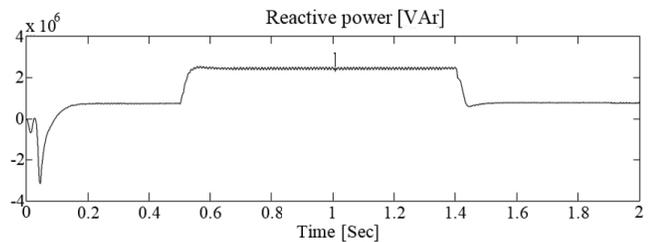
Fig. 12. Injected reactive power by the wind turbine during the fault

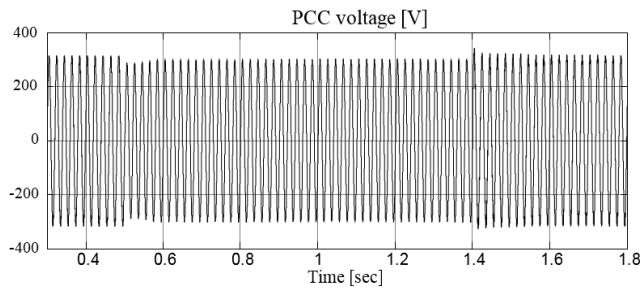
Fig. 13. Voltage drop improvement up to 70% at the PCC during the fault

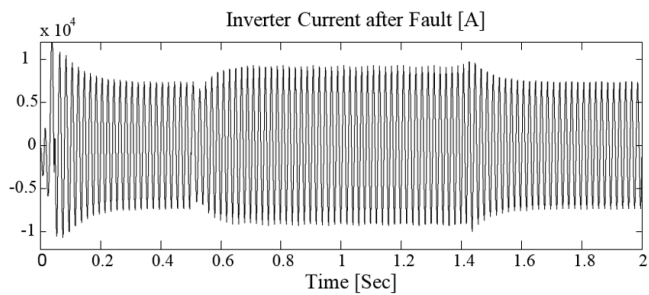
Fig. 14. Injected current from the wind turbine to the grid during the fault

## V. CONCLUSION

In this paper, the control process of the B2B converter in a wind turbine connected to a power grid is investigated in both normal and fault conditions. By applying the vector control method and the PWM strategy in the converter, the control objectives are achieved. The most important objectives are to compensate the voltage drop during fault time, as well as to track maximum available active power injected to the grid by the wind turbine. Assessing the system performance under the normal condition, and also when a three-phase fault occurs at the PCC indicates the satisfying performance of the B2B converter's controller in both conditions. In other words, while the controller in wind turbine system performs efficiently in the normal condition of the power grid, during a fault, the controller increases the injection of reactive power which results in the voltage drop improvement up to 70%.